\documentclass[fleqn,10pt]{wlscirep}
\usepackage[utf8]{inputenc}
\usepackage[T1]{fontenc}
\usepackage{comment}
\usepackage{orcidlink}

\usepackage{color}
\usepackage{bm}
\usepackage{graphicx}

\begin{document}
\title{Programmable Quantum Linear Interference with Pulse Shaping of Quantum Light}

\author[1,*]{Aruto Hosaka}
\author[1]{Masaya Tomita}
\author[2]{Yoshiaki Tsujimoto\orcidlink{0000-0002-7877-3797}}
\author[1]{Shintaro Niimura}
\author[1]{Akihito Omi}
\author[2]{Kentaro Wakui\orcidlink{0009-0002-6460-1676}}
\author[2]{Mikio Fujiwara}
\author[1,2]{Masahiro Takeoka}
\author[1]{Fumihiko Kannari}

\affil[1]{Department of Electronics and Electrical Engineering, Keio University, 3-14-1 Hiyoshi, Kohoku, Yokohama 223-8522, Japan}
\affil[2]{Advanced ICT Research Institute, National Institute of Information and Communications Technology (NICT), Koganei, Tokyo 184-8795, Japan}

\affil[*]{Hosaka.Aruto@cj.MitsubishiElectric.co.jp}

\begin{abstract}
In this paper, we propose a novel method for interfering frequency-multiplexed photonic quantum states without the use of optical nonlinear effects, and experimentally demonstrate this technique via frequency-domain Hong-Ou-Mandel (HOM) interference. By cascading the generation of quantum states onto arbitrary orthogonal modes, we can induce interference across any desired frequency mode. Following the generation of quantum states onto the frequency modes, performing measurements in independent frequency bands enables the realisation of a frequency-domain linear optical circuit analogous to linear interference in the spatial domain. We successfully demonstrated programmable quantum interference by controlling the spectral mode functions and measurement bases. Our method offers a new approach to harness the full potential of light's temporal-frequency degrees of freedom, providing a path towards scalable and programmable photonic quantum computing architectures without the need for optical nonlinearities or spatial-mode beam splitters.
\end{abstract}
\maketitle

Light possesses degrees of freedom in three distinct characteristics: spatial (angular frequency), frequency (time), and polarisation \cite{fabre2020modes}. Embedding and manipulating photonic quantum states within these modal degrees of freedom is one of the crucial tasks in achieving large-scale, programmable quantum computing. In previous research, numerous experiments in quantum optics have been conducted using the spatial degrees of freedom \cite{space1,space2,space3,space4,space5,space6}. Particularly, the development of photonic chips aimed at large-scale quantum information processing has been vigorous \cite{chip1,chip2,chip3,chip4,chip5}. This has led to the ability to achieve large-scale circuits with high precision, marking significant progress in the field.

In addition to spatial degrees of freedom, research on the utilisation of temporal-frequency degrees of freedom has also been progressing. To fully leverage the degrees of freedom in time and frequency, it is necessary to employ time-division multiplexing and frequency-division multiplexing, as is commonly done in the realm of classical communications \cite{WDM-TDM}. With respect to time multiplexing, several experiments have been conducted, demonstrating quantum optical experiments in large-scale temporal modes \cite{time1,time2,time3,time4,time5}.

In the frequency domain, researchers are exploring ways to generate and control quantum states \cite{WDM1,WDM2,WDM3,WDM4,WDM5,WDM6}. This presents certain challenges, as quantum states between different frequency modes do not interfere with each other when using passive devices like beam splitters. To overcome this, the active use of optical effects, such as sum-frequency mixing processes \cite{SFG1} or frequency modulators \cite{SFG2}, is being investigated.

In this paper, we propose a method for interfering frequency-multiplexed photonic quantum states without the use of optical nonlinear effects, and we experimentally demonstrate this technique via frequency-domain Hong-Ou-Mandel (HOM) interference \cite{HOM}. By repeating the generation of quantum states onto arbitrary orthogonal modes in a cascaded manner, we can induce interference across any desired frequency mode. Following the generation of quantum states onto the frequency modes, performing measurements in independent frequency bands enables the realization of a frequency-domain linear optical circuit that is analogous to linear interference in the spatial domain.

\begin{figure}[t]
\centering
\includegraphics[width=11.5cm]{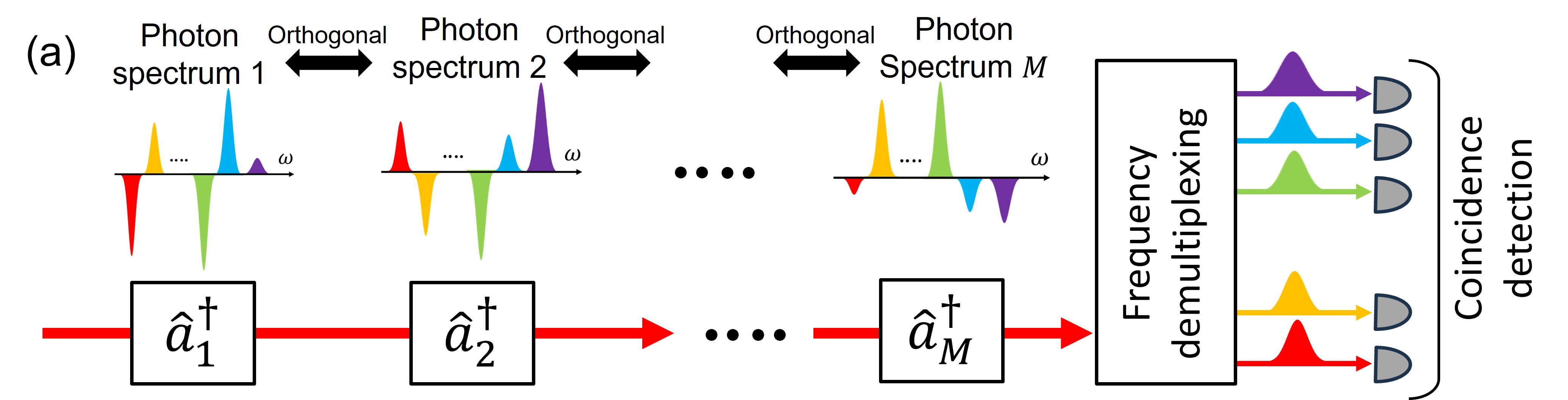}
\includegraphics[width=5.5cm]{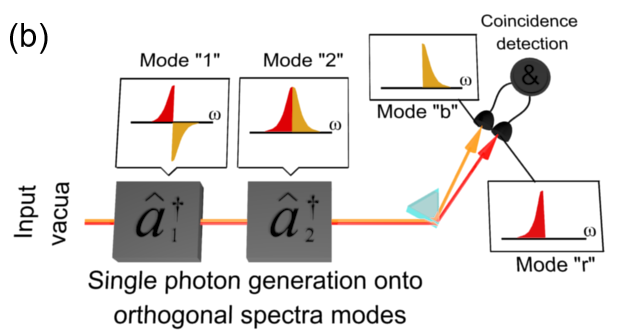}
\caption{Schematic description of a quantum linear interferometer without beam splitters. a, Schematic of generalized quantum interferometer. Each operations coherently generate quantum states onto the orthogonal vacuum modes. b, Frequency-domain HOM interference without beam splitters. Single photons are cascadingly generated on the orthogonal spectra but the same spatial mode, respectively. The generated photons are divided into independent spectral modes which is different from the generated modes. Destructive interference pattern is obtained when coincidence measurement is applied between two independent spectral modes.}
\label{fig1}
\end{figure}

Figure \ref{fig1} illustrates the schematic of our proposed frequency-domain photonic interferometer. Herein, we utilize $N$ independent frequency bins to generate $M$ photons (where $N>M$). Considering the incidence of $M$ photons into an $N$-mode linear interferometer, this mode mixing process can be expressed as $\hat{a}^\dagger_k = U_{k,1} \hat{b}^\dagger_1 + U_{k,2} \hat{b}^\dagger_2 +\dots+U_{k,N} \hat{b}^\dagger_N$. Here, $\hat{a}^\dagger_m$ and $\hat{b}^\dagger_n$ correspond to creation operators on the mixed modes and the frequency bins, respectively. $U$ corresponds to the unitary matrix representing the linear interference, satisfying the orthonormality condition $\sum_{n=1}^N U_{k,n} U_{k',n}=\delta_{k,k'}$, where $\delta_{k,k'}$ is the Kronecker delta. Considering this orthogonality, the commutation relations $[\hat{b}_n, \hat{b}^\dagger_{n'}]=\delta_{n,n'}$ imply that $[\hat{a}_m, \hat{a}^\dagger_{m'}]=\delta_{m,m'}$. This signifies that when photons are generated in a certain frequency mixed modes, if orthogonality is maintained among these modes, the photon generation processes are completely independent of each other.

Figure \ref{fig1}b is the schematic view of the frequency-domain HOM experiment as the simplest case of our proposed method. Firstly, single photons are generated onto the frequency mode "1" and "2". The mode function of the frequency mode "2" shows a Gaussian spectrum, whereas the mode function of the frequency mode "1" has the spectrum whose long wavelength component is $\pi$ phase shifted relative to that of the mode function "2".
Since modes "1" and "2" are orthogonal, $\hat{a}^\dagger_2$ does not affect the mode "2" in which photons are being generated by $\hat{a}^\dagger_1$. Here $\hat{a}^\dagger_1 = (\hat{b}^\dagger_1 - \hat{b}^\dagger_2)/\sqrt{2}$ and $\hat{a}^\dagger_2 = (\hat{b}^\dagger_1 + \hat{b}^\dagger_2)/\sqrt{2}$. When viewed in the basis of frequency bins, we have $\hat{a}^\dagger_1 \hat{a}^\dagger_2 = (\hat{b}^{\dagger2}_1-\hat{b}^{\dagger2}_2)/2$, which leads to Hong-Ou-Mandel (HOM) interference. Thus, as shown in Fig. \ref{fig1}b, by choosing the final measurement basis to be the frequency bins, we can observe a reduction in the coincidence count rate.

\begin{figure}[ht]
\centering
\includegraphics[width=12cm]{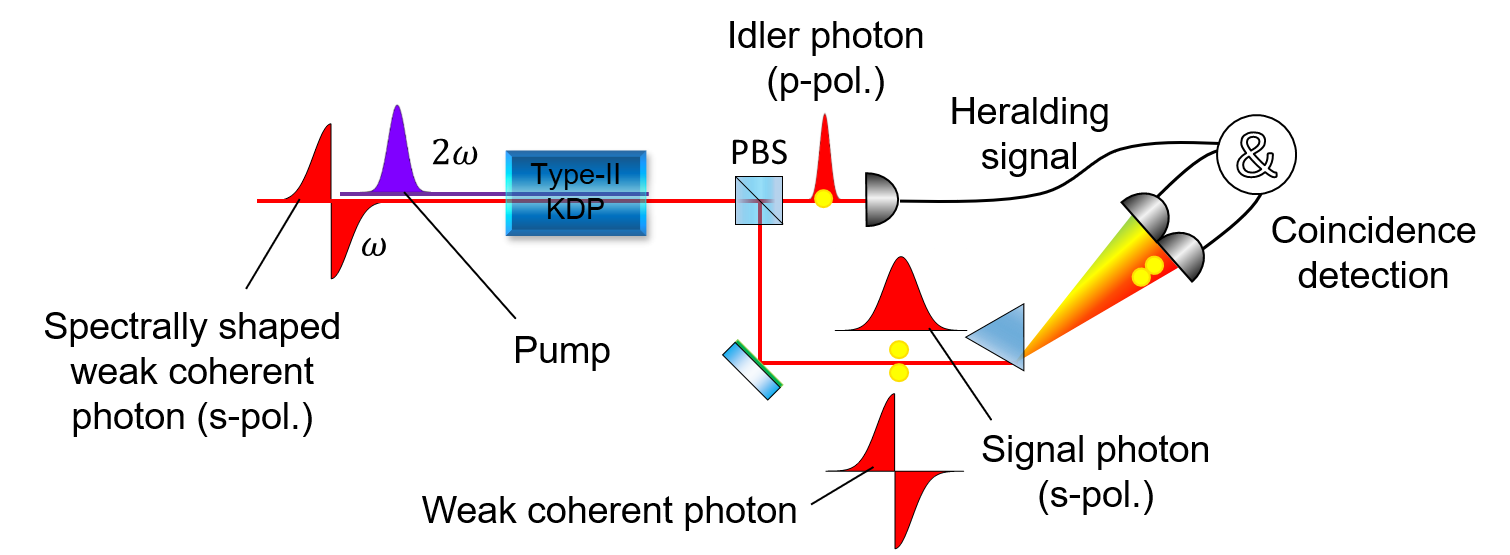}
\caption{Schematic view of the experimental setup.}
\label{fig2}
\end{figure}
Figure 2 shows the schematic view of our experimental setup. In the experiment, we obtained HOM interference between a weak coherent state and a heralded single photon state \cite{HOM_LO,HOM_JIN}. As a light source, we employed a modelocked Ti:sapphire laser whose center wavelength is 830 nm.
The pules width and the repetition rate were 100 fs and 80 MHz. The modelocked pulses were frequency-doubled with a 500-${\rm\mu m}$ Type-I BBO crystal, and the remained fundamental pulses were divided by a color filter. The fundamental pulses were attenuated to a single photon level by a neutral density filter and spectrally shaped by a 4-$f$ pulse shaper based on the method introduced by Frumker and Silberberg\cite{LCOS}. Correspondigly, we generated a weak coherent state $\left| \alpha \right>_1$ on the frequency mode "1". The shaped weak coherent pulses and the second-harmonic pulses were recombined with a color filter and synchronously incident to a type-II KDP crystal. In the crystal, a signal and idler photon pair was generated through a spontaneous parametric down conversion. Here, the frequency modes of the generated signal photon and incident weak coherent state are orthogonal to each other, so no induced emission occurs and the coherent state just passes through the crystal. The heralded single photons $\left| 1 \right>_2$ on the mode "2" were successfully obtained by using the idler photon as the heralding signal. The orthogonality between $\left| \alpha \right>_1$ and $\left| 1 \right>_2$ was confirmed by the visibility of the spatial-domain HOM interference (Detailed in Supplementary Information Sec. 6). Dividing the frequency components of the obtained coherent state and the single photon with a grating-based wavelength separator, we measured the coincidence count rate between long- and short-wavelength components and verified the HOM interference.

\begin{figure}[ht]
\centering
\includegraphics[width=12cm]{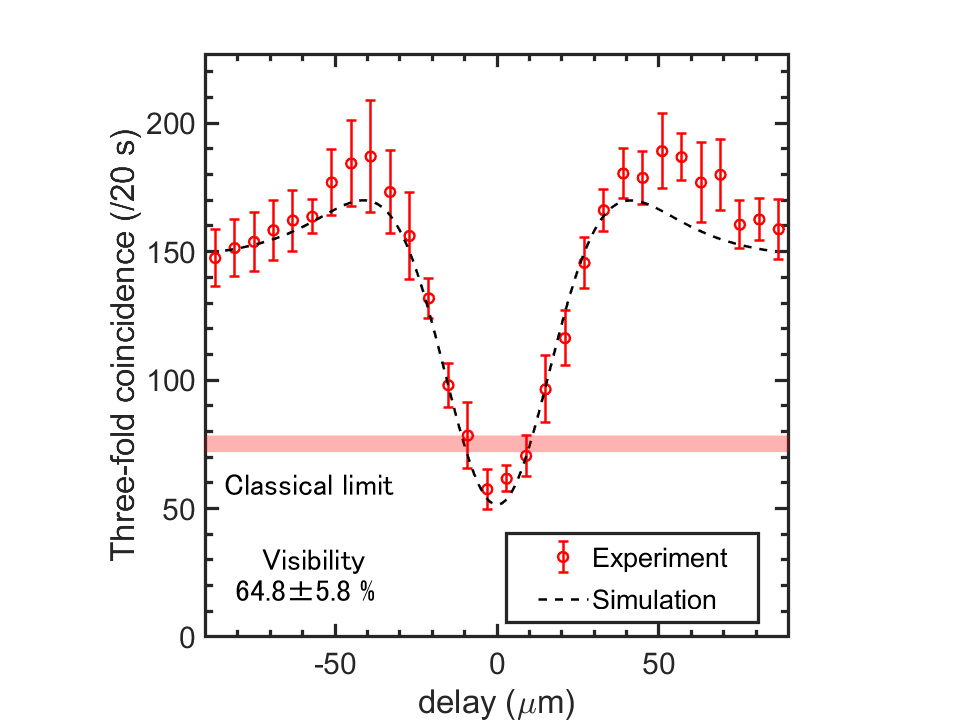}
\caption{The HOM interference without beam splitters. Red circles indicate experimental coincidence count rate. Black dashed line indicates numerical simulation results. Red filled area represents the classical limit of the HOM dip.}
\label{fig3}
\end{figure}

Figure 3 shows the results of the HOM interference without beam splitters. The visibility of the HOM dip was $64.8 \pm 5.8\% $ which is beyond the classical limit of $50 \%$.  We observed an increasing of the three coincidence count rate at the non-zero delay point (around $\pm 40$ $\mathrm{\mu m}$). When the delay is not zero, the modes of the coherent state and heralded single photon are no longer orthogonal to each other. Therefore, the induced emission causes an optical parametric amplification. Such inorthogonality at the non-zero delay point can be seen in the spatial-domain HOM interference between the shaped coherent state and the heralded single photon (Figure 10 of Sec. 6 in Supplementary information). In the current experiment, the visibility is deteriorated by the spectral inpurity of the heralded single photon \cite{KDP} and multi-photon effect (Discussed in Supplementary information Sec. 5).

Next, we demonstrated the programmability of our proposed scheme. As shown in Fig. 4, we divided the heralded single photon's spectrum into two components at the intensity ratio of $T:R$. The weak coherent state, on the other hand, is spectrally shaped so that the spectral intensity is divided at the ratio of $R:T$ and a $\pi$ phase shift is added to the long-wavelength component. In this way, we can achieve any two-mode mixing without beam splitters.

\begin{figure}[ht]
\centering
\includegraphics[width=8cm]{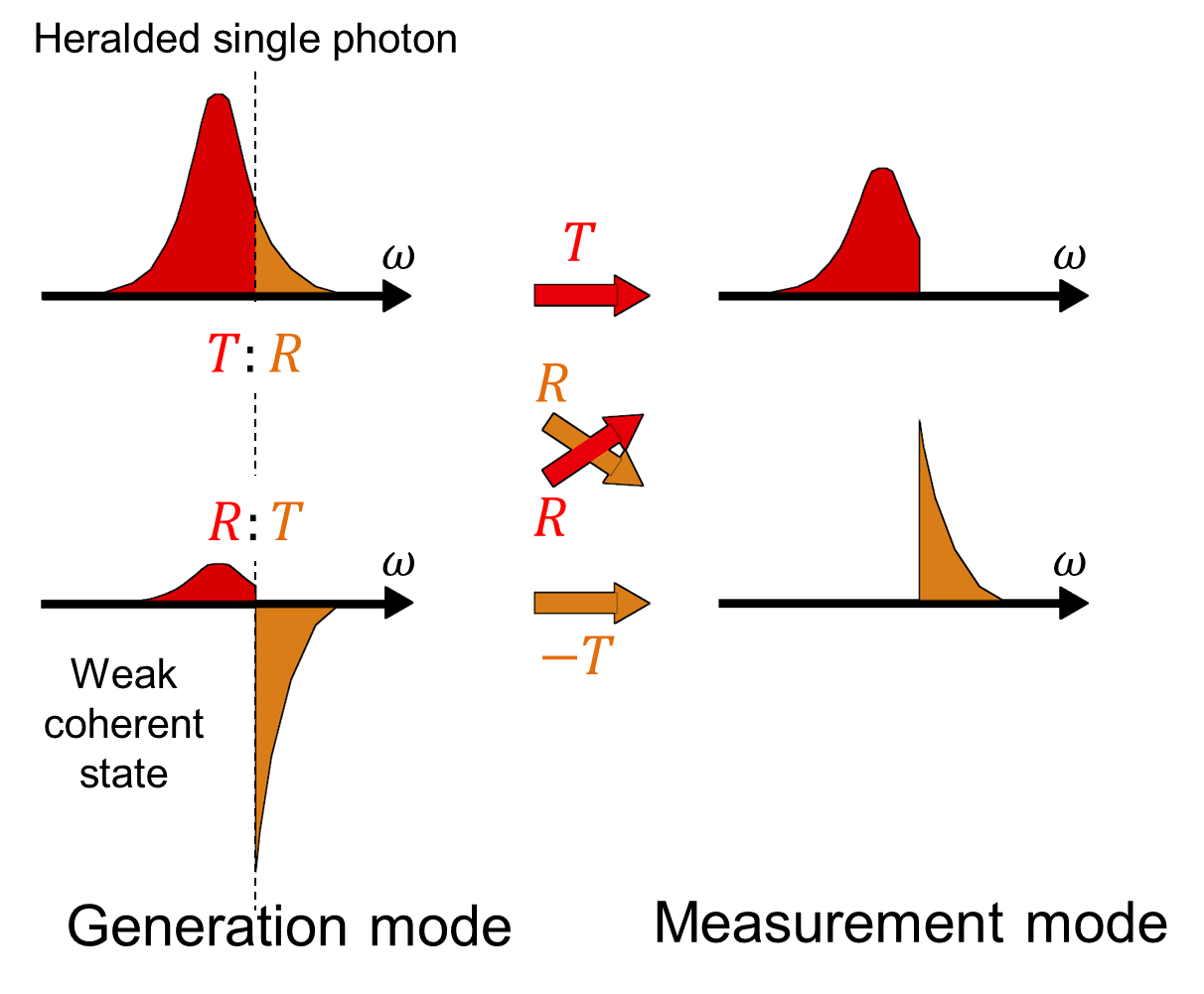}
\caption{Principle of the programmable quantum interference without beam splitters.}
\label{fig4}
\end{figure}

\begin{figure}[ht]
\centering
\includegraphics[width=16cm]{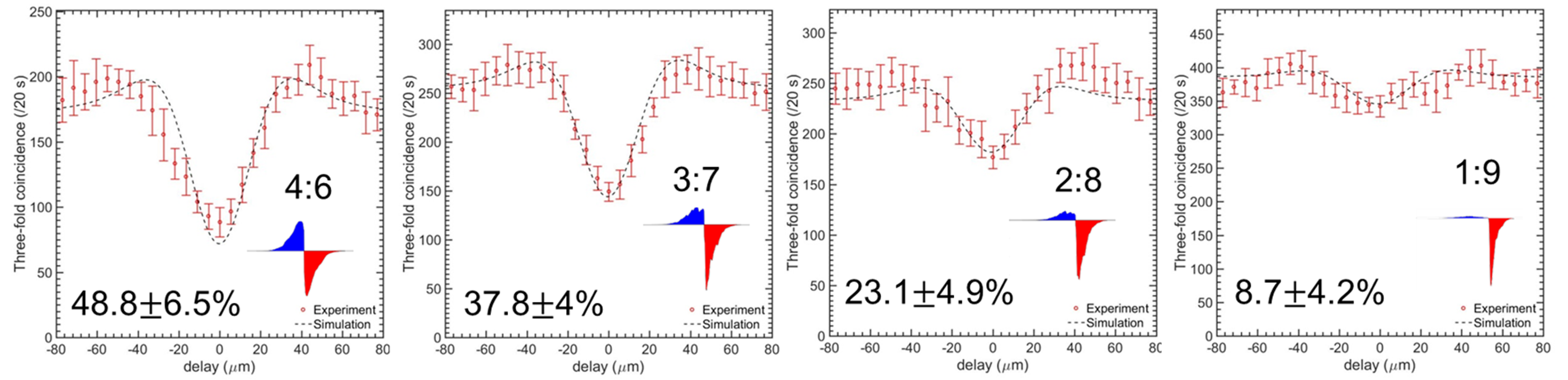}
\caption{Experimental results of the programmable quantum interference without beam splitters.}
\label{fig5}
\end{figure}

Figure 5 shows the experimental results of the programmable quantum interference without beam splitters. We experimentally demonstrated 4:6, 3:7, 2:8 and 1:9 ratios of the two-mode mixing. Likewise for the 50:50 ratio two mode mixing, the orthogonalities between the heralded single photon's spectral mode and all spectral modes of the spectrally shaped weak coherent pulses were verified by the spatial domain HOM interference. As shown in Fig. 5, we successfully  controlled the visibilities of the two photon quantum interference. Currently, the visibility is limited by high-order Schmidt modes and multiphoton probability, but these factors can be resolved by dispersion control of the pump pulse \cite{BJS}.

In conclusion, we proposed a method to realize a programmable frequency-domain linear interference without nonlinear mode mixing. By cascadely generating quantum states onto the designed modes and measuring independent frequency bins, we demonstrated that the quantum interference can also be fully programmed. In our method, the most difficult part is to generate quantum states onto the designed mode. Ansari {\it et al.} and Omi {{\it et al.}} proposed the method to generate heralded single photon in the shaped spectral mode by dispersion-engineered SPDC \cite{PSphoton}. This technique can be also applied to squeezed state and photon number state generation by replacing the heralding detectors by homodyne detectors and photon-number-resolved detector, respectively. Therefore, our method is directly applicable even to the scatter-shot boson sampling \cite{GBS1}, Gaussian boson sampling \cite{GBS2} and continuous-variable cluster state generation \cite{MBQC} as well as boson sampling with heralded single photons. Our proposed method for frequency-domain interference of photonic quantum states offers a novel approach to harnessing the full potential of light's degrees of freedom, paving the way for the advancement of scalable and programmable photonic quantum computing architectures.

\section*{Acknowledgements}
This work is supported by JST CREST (Grant No. JPMJCR1772) and JSPS KAKENHI (Grant Nos. 17K05082 and 19H05156).

\section*{Author contributions}
A.H. conceived the idea, designed and performed the experiments, and analyzed the data. M.T. performed the experiments and data analysis. Y.T. discussed the experimental design and provided guidance. S.N. and A.O. built the experimental setup. K.W. and M.F. provided technical support for the experiments. M.T. contributed to the theoretical aspects of the work. F.K. supervised the project and participated in all discussions related to the experiments. All authors contributed to the writing of the manuscript.

\section*{Competing interests}
The authors declare that they have no competing interests.

\section*{Author Information}
Correspondence and requests for materials should be addressed to A.H.~(email: Hosaka.Aruto@cj.MitsubishiElectric.co.jp).

\bibliography{nature}

\clearpage
\part*{Supplementary information}

\section{Heralded single photon}
As a single photon source, we used a heralded single photon generated via spontaneous parametric down conversion. To generate a spectrally pure photon pair at 830-nm, a 15-mm Type-II KDP crystal was pumped by a 60-fs pulse whose center wavelength was 415 nm []. We experimentally obtained the joint spectral intensity (JSI) as shown in fig. 6(a). Figure 6(b) shows the singular value calculated from the JSI. The purity of the photon pair calculated from JSI was 0.91 assuming the pump pulse was Fourier transform limited, and the corresponding Schmidt number was 1.10. The first Schmidt mode function are shown in Fig. 6(c) and (d).

\begin{figure}[ht]
\centering
\includegraphics[width=12cm]{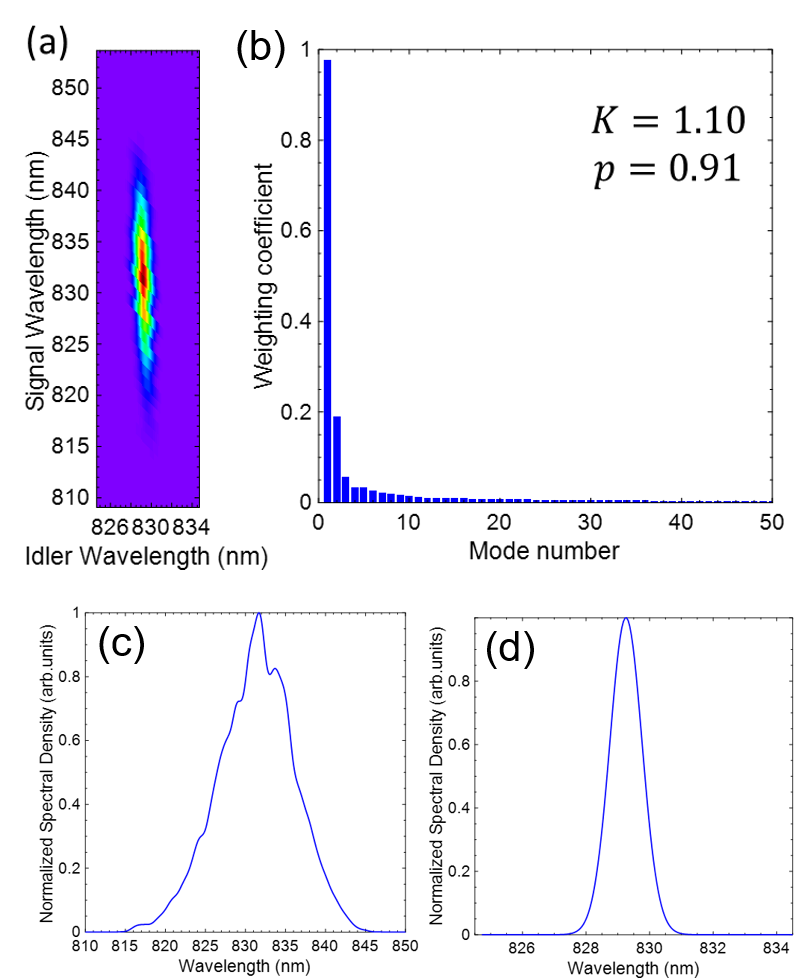}
\caption{Characteristics of the heralded single photon generated by the 15-mm type-II KDP crystal.}
\label{fig6}
\end{figure}

The $g^{(2)}$ parameter of the photon pair was $0.043\pm0.001$ when the HOM dip shown in Fig. 3 was obtained. The photon per pulse of the coherent pulse was ????.

\section{Pulse shaper}
We used $1440\times1050$-pixel 2-D LCOS (Santec, SLM-100) for the amplitude and phase modulation [????]. The input pulse was divided into spectral components by a grating of 1200 lines/mm. Then, the beam was collimated by a cylindrical lens with the focal length of 200 mm. The 2-D LCOS was located at the Fourier plane, and phase modulation was applied to the each spectral components. The amplitude and phase modulated first-order diffracted beam was used as a photon source for the HOM inerference.  The measured resolution of the pulse shaper was ~0.1 nm (FWHM).

To match the bandwidth of the shaped coherent photons to the signal photons generated via spontaneous parametric down conversion (SPDC), we applied amplitude modulation to the coherent photons so that its bandwidth was broadened (Fig. 7). The bandwidth was broadened form 8.2 nm to 11.7 nm by the gausian amplitude modulation. Furthermore, we added 3000-fs$^2$ positive chirp to the coherent pulse so that the visibility of the HOM interference was improved.

\begin{figure}[ht]
\centering
\includegraphics[width=12cm]{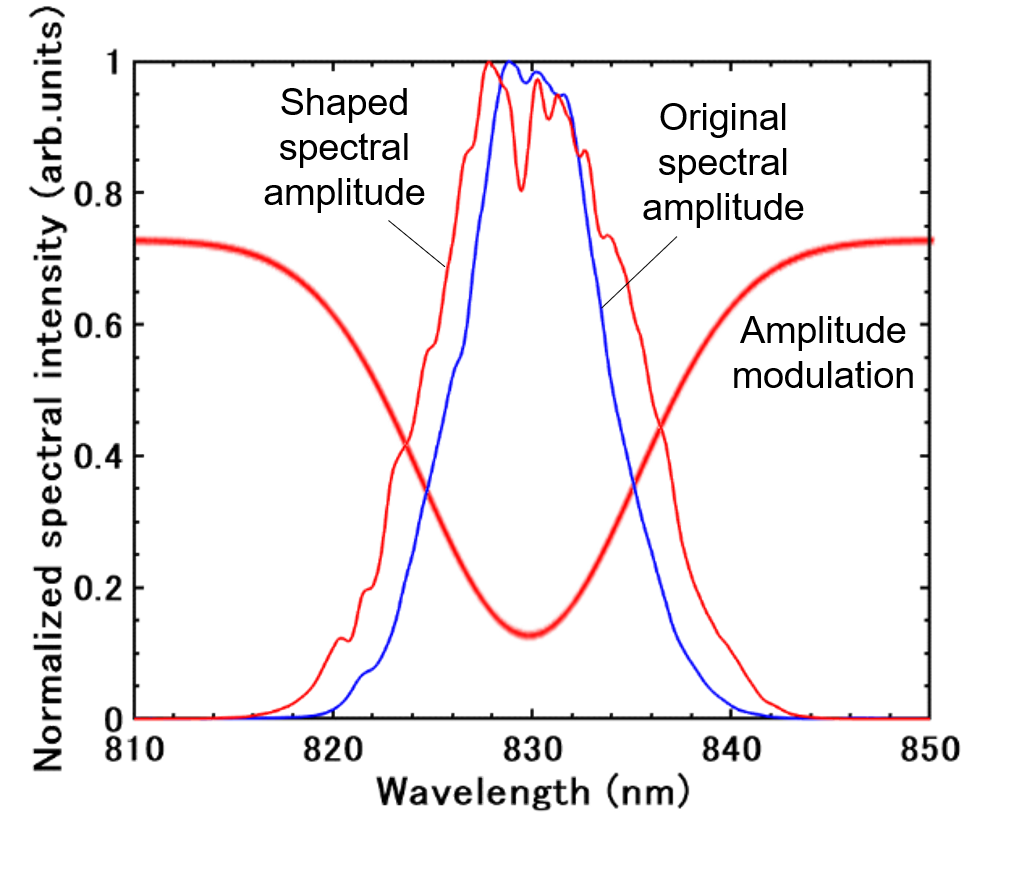}
\caption{Spectrum of the weak coherent pulse. The red and blue curve show the shaped and the original spectra, respectively. The red bold curve shows amplitude modulation applied to the original spectrum to broaden its spectral width.}
\label{fig7}
\end{figure}

\section{Wavelength separator}
The wavelength separator divided the incident pulse into long- and short-wavelength component. Figure 8a shows the schematic view of the wavelength separator. The pulse delivered through the single mode-fiber was incident to the 1200-groove/mm grating. Then, the pulse dispersed  into spectral components were collimated by a lens with the focal length of 300 mm. At the Fourier plane, the beam was divided into the two components by the D-shaped mirror. The dividing point could be adjusted by moving this mirror. The divided beam were incident to the multimode fiber with the core diameter of 50 $\mathrm{\mu m}$, respectively. The spectra divided at the center wavelength is shown in Fig. 8b, and it can be seen the spectra was almost uniformly divided.

\begin{figure}[ht]
\centering
\includegraphics[width=16cm]{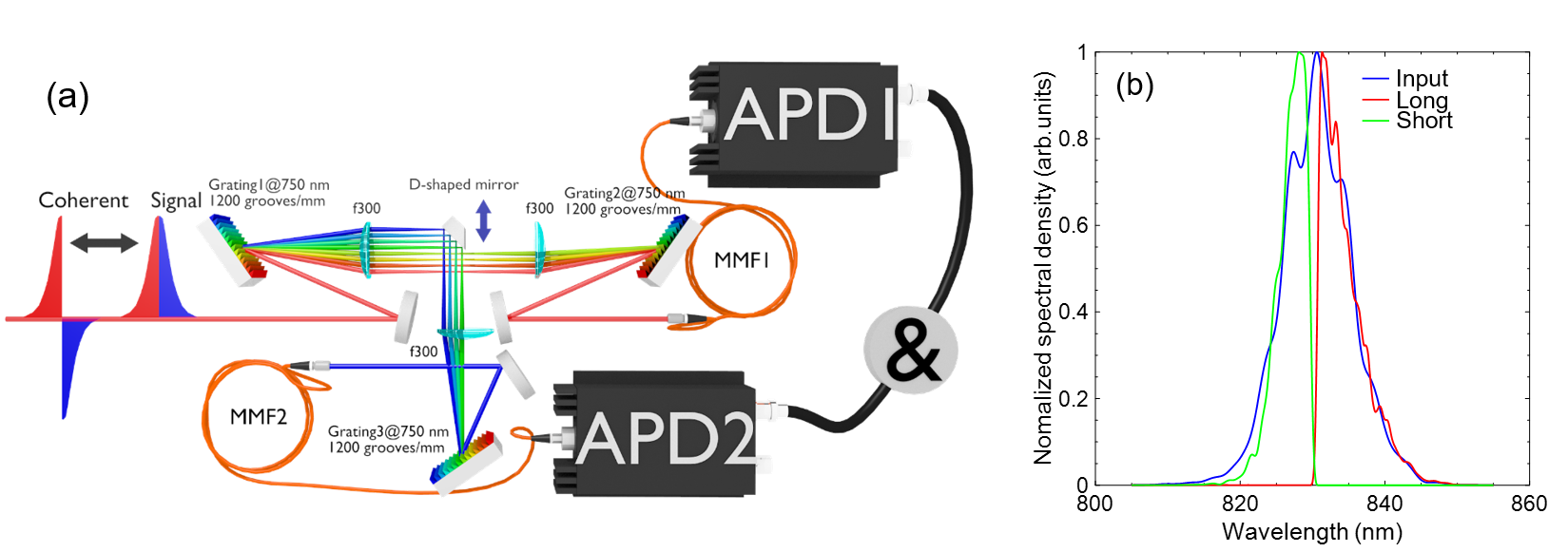}
\caption{(a) The wavelength separator used in the experiment. (b) The input spectrum and the spectra of the divided long- and short-wavelength components are drawn as the blue, red and green curves, respectively..}
\label{fig8}
\end{figure}

\section{Induced emission}
When the two photons do not have orthogonal modes, the induced emission increase the coincidence count rate. The induced emission caused by the nonlinear crystal can be expressed as

\begin{equation}
 \hat{S}_{s,i}\left| \alpha\right>  \approx (1+\gamma \hat{a}_i^\dagger \hat{a}_s^\dagger)( 1+ \alpha_1 \hat{a}_s^ \dagger) \left| 0_s \right>\left| 0_i \right> = \sqrt{2}\gamma\alpha \left| 2_s \right>\left| 1_i \right> + \gamma\left| 1_s \right>\left| 1_i \right> + \alpha\left| 1_s \right>\left| 0_i \right> + \left| 0_s \right>\left| 0_i \right>,
\end{equation}

where $\hat{S}_{s,i}$ is two-mode squeezing operator with nonlinear gain $\gamma$. $\alpha$ is amplitude of the weak coherent state. Here, the three coincidence count probability $|\gamma\alpha|^2$. When the weak coherent pulse and the down converted photon are not overlapped, the three is $|\gamma\alpha|^2/2$. Therefore, it is found that the coincidence probability becomes twice due to the induced emission.

Figure 9 shows the results of the HOM interference (red circles) and the induced parametric amplification (blue circles). In this experiment of the induced emission, the $\pi$-phase shift for orthogonalizing was not applied to the weak coherent pulse. As a result, the single photons launched into the KDP crystal caused induced emission, and photon pair generation rate was increased. As we derived, although the photon pair generation rate should be twice, it was 1.64 times as shown in Fig. 9 due to the high-order Schmidt mode discussed in the Sec. 5 in this supplementary information. The dashed line in Fig. 9 indicates the simulation result of the induced emission.

\begin{figure}[ht]
\centering
\includegraphics[width=12cm]{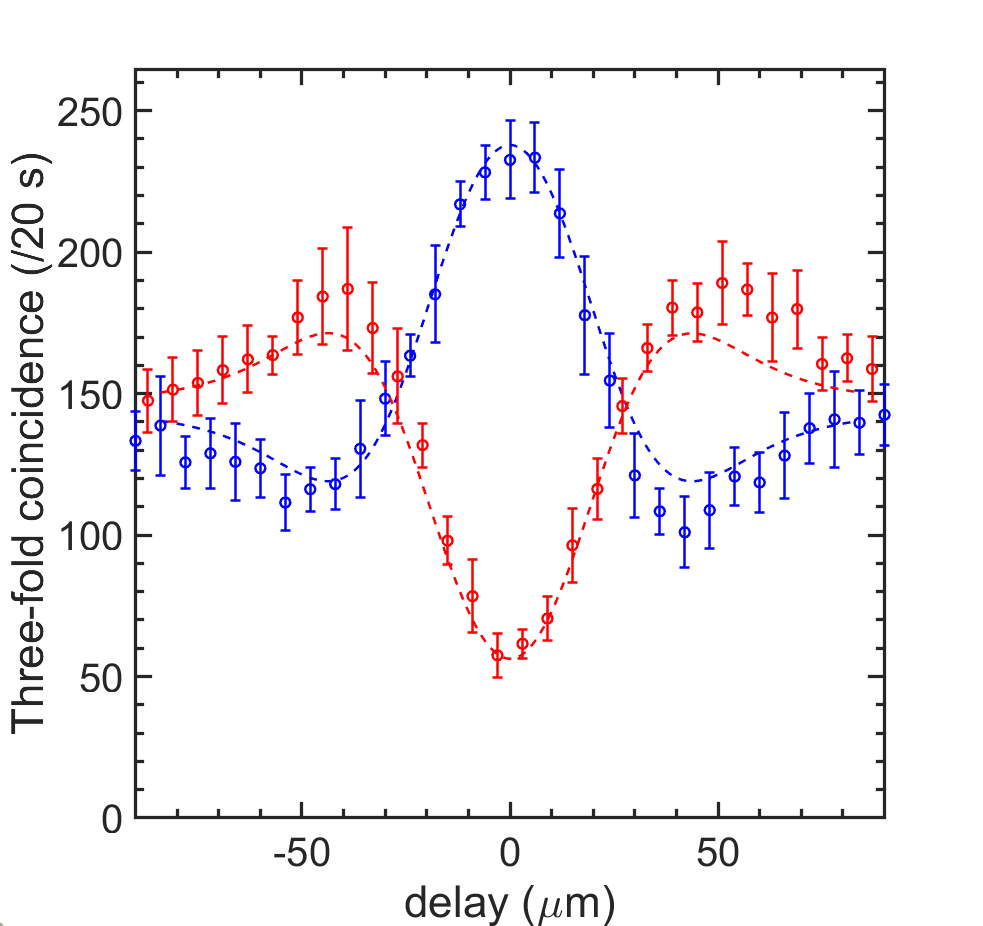}
\caption{Experimental Three-fold coincidence count rates of the HOM interference (red circles) and the induced emission (red circles). The dashed lines are simulation results based on the calculation described in sec. 5.}
\label{fig9}
\end{figure}

\section{Numerical simulation of HOM interference without mode mixers}
In this section, we describe the method to numerically simulate the experiment of the HOM interference. To fully analyze SPDC process, it is necessary to take high-order Schmidt mode into consideration. First, we define spectral Schmidt modes. $n$th Schmidt mode function of the signal and idler photons are defined as $\psi_{s,n}(\omega)$ and $\psi_{i,n}(\omega)$, respectively. $\pi$ phase shifted mode function of spectrally shaped coherent state is defined as $g_1(\omega)$. Gaussian mode function orthogonal to $g_1(\omega)$ is $g_2(\omega)$.

A weak coherent state generated on the mode $g_1(\omega_s)$ and $g_2(\omega_s)$  can be approximated as

\begin{equation}
  \left| \alpha_1\right>  \approx  ( 1+ \alpha_1 \hat{a}_1^ \dagger + \frac{\alpha_1^2}{2} \hat{a}_1^{\dagger^2}) \left| 0_1 \right>,
\end{equation}
\begin{equation}
  \left| \alpha_2\right>  \approx  ( 1+ \alpha_2 \hat{a}_2^ \dagger + \frac{\alpha_2^2}{2} \hat{a}_2^{\dagger^2}) \left| 0_2 \right>,
\end{equation}
where $\hat{a}_1^ \dagger$ and $\hat{a}_2^ \dagger$ is creation operators which acts on the modes $g_1(\omega_s)$ and $g_2(\omega_s)$, respectively. $\omega_s$ is angular frequency, and $\alpha_1$ is the amplitude of the coherent state which is sufficiently small to approximate.

We also define and approximate the two-mode squeezing operation in the KDP crystal in the multimode picture. We have to describe the spontaneous parametric down conversion in multimode  expression, and the Hamiltonian can be expressed as 
\begin{equation}
H = \gamma \int\int \theta(\omega_s, \omega_i) f(\omega_s+\omega_i) \hat{a}_s^\dagger(\omega_s) \hat{a}_i(\omega_i)^\dagger d\omega_i d\omega_s + \rm{H.C.}
\end{equation}
$\gamma$, $\theta(\omega_s, \omega_i)$ and $f(\omega_s+\omega_i)$ are nonlinear coefficient, phase matching conditions and pump spectral amplitude, respectively. $\hat{a}_s^\dagger(\omega_s)$ and $\hat{a}_i(\omega_i)^\dagger$ are creation operators acting on the signal and idler fields, respectively. It is known that this Hamiltonian can be rewritten into the simple uncorrelated modal structure by Schmidt decomposition
\begin{equation}
H = \sum_{n=0}^{\infty} \gamma_n \hat{a}_{s,n}^\dagger \hat{a}_{i,n}^\dagger + \rm{H.C.}
\end{equation}

$\hat{a}_{s,n}^\dagger$ and $\hat{a}_{i,n}^\dagger$ are the creation operator acting on the $n$th Schmidt modes
\begin{eqnarray}
  \hat{S}_n(\gamma_n)  & \approx & 1 + \gamma _n \hat{a}^\dagger_{s,n}\hat{a}^\dagger_{i,n} + \frac{\gamma _n^2}{2} (\hat{a}^\dagger_{n,s}\hat{a}^\dagger_{n,i})^2 \nonumber \\
& = & 1 + \gamma _n (c_{1,n} \hat{a}^\dagger_1 + c_{2,n} \hat{a}^\dagger_2 + c_{e,n} \hat{a}^\dagger_{e,n}) \hat{a}^\dagger_{i,n} + \frac{\gamma _n^2}{2} (c_{1,n} \hat{a}^\dagger_1 + c_{2,n} \hat{a}^\dagger_2 + c_{e,n} \hat{a}^\dagger_{e,n}) ^2 \hat{a}^{\dagger^2}_{i,n} 
\end{eqnarray}
HOM interference in the frequency domain can be written as
\begin{eqnarray}
  \hat{S}_n(\gamma_n) \left| \alpha_1\right> \left| 0_2\right> \left| 0_{e,n}\right> \left| 0_{i,n}\right> \approx   \left[ 1 + \gamma _n (c_{1,n} \hat{a}^\dagger_1 + c_{2,n} \hat{a}^\dagger_2 + c_{e,n} \hat{a}^\dagger_{e,n}) \hat{a}^\dagger_{i,n}  + \frac{\gamma _n^2}{2} (c_{1,n} \hat{a}^\dagger_1 + c_{2,n} \hat{a}^\dagger_2 + c_{e,n} \hat{a}^\dagger_{e,n}) ^2 \hat{a}^{\dagger^2}_{i,n} \right] \nonumber \\
  ( 1+ \alpha_1 \hat{a}_1^ \dagger + \frac{\alpha_1^2}{2} \hat{a}_1^{\dagger^2}) \left| 0_1\right> \left| 0_2\right> \left| 0_{e,n}\right> \left| 0_{i,n}\right>
\end{eqnarray}

By applying mode mixing operator $\hat{U}_{1,2}$ and $\hat{U}_{e,o,n}$ acting on $\{\left | \psi_1 \right >, \left | \psi_2 \right >\}$ and $\{\left | \psi_{e,n} \right >, \left | \psi_{o,n} \right >\}$, respectively, where $\left | \psi_{o,n} \right >$ is the orthogonal and complementary mode to $\left | \psi_{e,n} \right >$, we obtained final quantum states. We calculated all the combinations and acquired the theoretical curve shown in Fig. 3 and Fig. 9 based on the experimentally obtained parameter.

\section{Verification of orthogonality}
To check the orthogonality between the spectrally shaped weak coherent state and the heralded single photon, we experimentally obtained the conventional spatial-domain HOM interference between them. If the distinguishable two photons are interfered with a beam splitter, the HOM interference are not obtained. We prepared a spectrally shaped photon and a heralded single photon in the same way written in the main text. These photons interfered each other with a 50:50 fiber beam splitter. The results of the spatial-domain HOM interferences are shown in Fig. 10. As shown in Fig. 10, we observed the elimination of the HOM dip at the point of the zero delay, and the orthogonality was verified in all the waveforms.

\begin{figure}[ht]
\centering
\includegraphics[width=16cm]{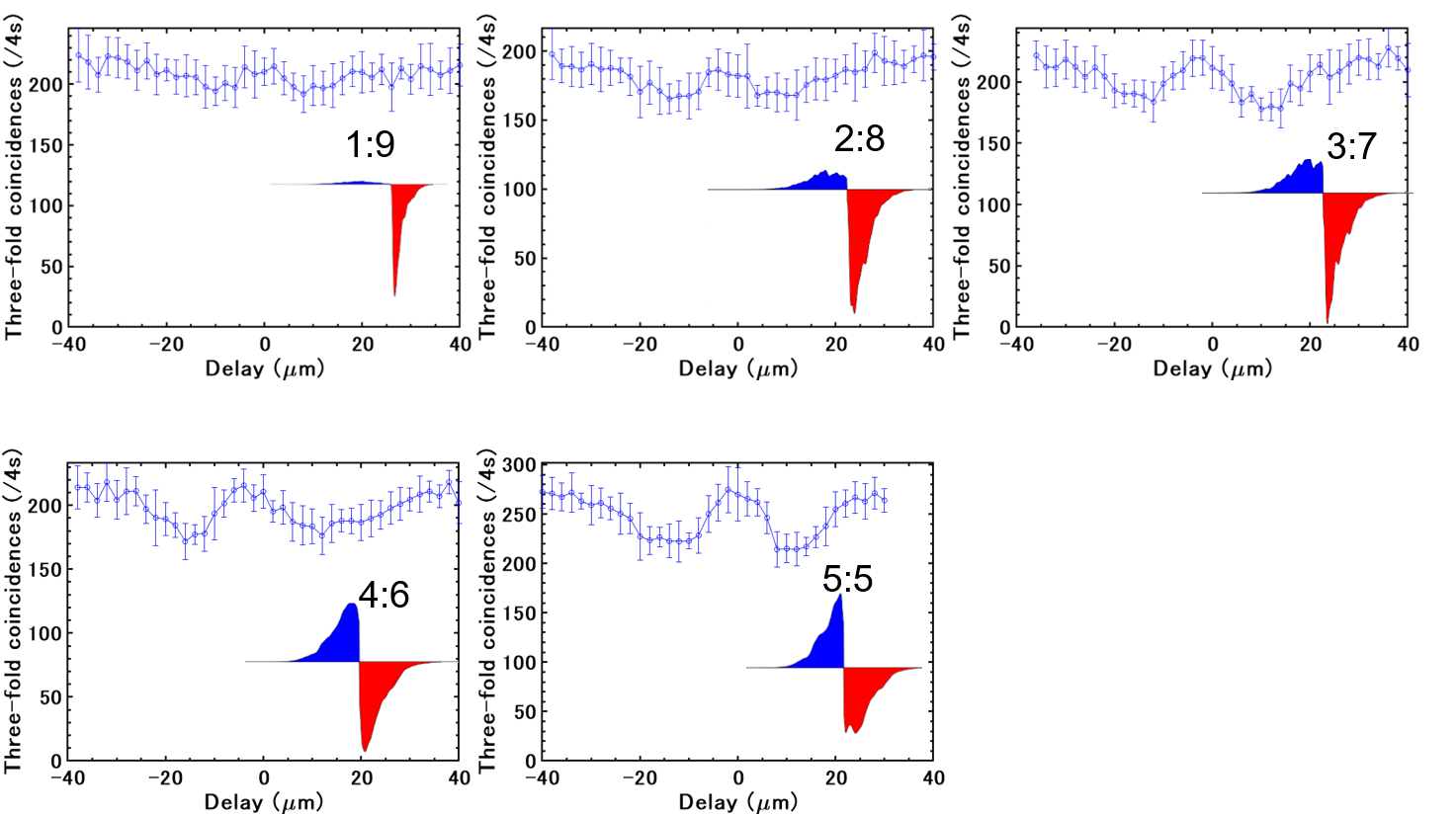}
\caption{Verification of the orthogonality between spectrally shaped pulses used in the experiment and Gaussian-like heralded single photon.}
\label{fig10}
\end{figure}

\end{document}